\documentclass[final,5p,times,twocolumn]{elsarticle}
 \biboptions{comma,sort&compress}
\usepackage{graphicx}
\usepackage{here}
\usepackage[dvips]{epsfig}
\usepackage{cuted}
\def\beq{\begin{equation}}
\def\eeq{\end{equation}}
\def\bea{\begin{eqnarray}}
\def\eea{\end{eqnarray}}
\def\bq{\begin{quote}}
\def\eq{\end{quote}}

\def\nnb{\nonumber}
\def\ga{\left(}
\def\dr{\right)}

\def\be{\beta}
\def\dl{\rm L_2}

												\def\nn{\nonumber}				
\def\nnb{\nonumber}
\def\la{\langle}
\def\ra{\rangle}
\def\nin{\noindent}
\def\ba{\vspace*{-0.2cm}\begin{array}}
\def\ea{\end{array}\vspace*{-0.2cm}}

\def\b{$\bullet~$}
\def\al{\alpha}
\def\als{\alpha_s}
\def\as{\ga\frac{\bar{\alpha_s}}{\pi}\dr}

\def\gg2{ \la\alpha_s G^2 \ra}
\def\gg3{g^3f_{abc}\la G^aG^bG^c \ra}
\def\ggg4{\la\als^2G^4\ra}

\journal{Elsevier}

\begin{document}
\begin{frontmatter}

\title{$\overline{m}_c$ and $\overline{m}_b$ from  $M_{B_c}$ and improved estimate of $f_{B_c}$
and $f_{B_c(2S)}$
} 
 \author[label2]{Stephan Narison}
\address[label2]{Laboratoire
Particules et Univers de Montpellier, CNRS-IN2P3, 
Case 070, Place Eug\`ene
Bataillon, 34095 - Montpellier, France.}
   \ead{snarison@yahoo.fr}

\markright{$\overline{m}_c$ and $\overline{m}_b$ from  $M_{B_c}$  and improved estimate of $f_{B_c}$}
\begin{abstract}
\noindent
We extract ({\it for the first time}) the correlated values of the running masses $\overline{m}_c$ and 
$\overline{m}_b$ from  $M_{B_c}$ using QCD Laplace sum rules (LSR) within stability criteria where pertubative (PT) expressions at N2LO and non-perturbative (NP) gluon condensates at LO are included. Allowing the values of  $\overline{m}_{c,b}(\overline{m}_{c,b})$ to move inside the enlarged range of recent estimates from charmonium and bottomium sum  rules (Table\,\ref{tab:param}) obtained using similar stability criteria, we deduce : $\overline{m}_c(\overline{m}_c)=1286(16)$ MeV and $\overline{m}_b(\overline{m}_b)=4202(8)$ MeV. Combined with previous estimates (Table\,\ref{tab:hmass}), we deduce a tentative QCD Spectral Sum Rules (QSSR) average :  $\overline{m}_c(\overline{m}_c)=1266(6)$\,MeV and $\overline{m}_b(\overline{m}_b)=4197(8)$\,MeV where the errors come from the precise determinations from $J/\psi$ and $\Upsilon$ sum rules.  As a result, we present an  improved prediction of $f_{B_c}=371(17)$\,MeV and the tentative upper bound  $f_{B_c(2S)}\leq 139(6)$\,MeV, which are useful for a further analysis of $B_c$-decays. 
\end{abstract}
\end{frontmatter}
 \section{Introduction}
Extractions of the perturbative (quark masses, $\alpha_s$) and non-perturbative quark and gluon condensates QCD parameters are very important as they will serve as inputs in different phenomenological applications of the (non)-standard model. 
Lattice calculations are an useful tool for a such project but alternative analytical approaches based on QCD first principles (Chiral perturbation, effective theory and QCD spectral sum rules (QSSR)) are useful complement and independent check of the previous numerical simulations as they give insights for a better understanding of the (non)-perturbative phenomena inside the hadron ``black box".

In this note\,\footnote{Some preliminary results of this work has been presented in\,\cite{SN19}.}, we shall use the Laplace version\,\cite{SVZa,SVZb,BELLa,BELLb,BECCHI,SNR}  of QSSR introduced by Shifman-Vainshtein-Zakharov (SVZ)\,\cite{SVZa,SVZb,ZAKA,BERTa,SNB1,SNB2,SNB3,SNB4,SNREV15,IOFFEb,RRY,DERAF,YNDB,PASC,DOSCH,COL} for a new extraction of the running quark masses $\overline{m}_c$ and $\overline{m}_b$ from the $B_c(0^{-+})$-meson mass which we shall use for improving the prediction on its decay constant $f_{B_c}$ done previously using similar approaches in\,\cite{SNBc,COL2,BAGAN,CHABAB,BAKER,SNp15}. 

\section{The QCD Laplace sum rules}
\subsection*{\b The QCD interpolating current}
We shall be concerned with the following QCD interpolating current:
\beq
\la 0|J_5(x)\vert P\ra= f_P M_P^2~:~J_5(x)\equiv (m_c+m_b)\bar c(i\gamma_5)
b~,
\label{eq:fp}
\label{eq:current}
\eeq
where: 
 $J_5(x) $  is the local heavy-light pseudoscalar current;   $m_{c,b}$ are renormalized mass of the QCD Lagrangian; $f_P$  is the decay constant related to  the leptonic widths $\Gamma [P\to l^+\nu_l]$ and normalised as $f_\pi=132$ MeV.
\subsection*{\b Form of the sum rules}
We shall work with the  Finite Energy version of the QCD Laplace sum rules (LSR) and their ratios :
\beq
{\cal L}^c_n(\tau,\mu)=\int_{(m_c+m_b)^2}^{t_c}\hspace*{-1cm}dt~t^n~e^{-t\tau}\frac{1}{\pi} \mbox{Im}~\psi_5(t,\mu)~,~~~
 {\cal R}^c_n(\tau)=\frac{{\cal L}^c_{n+1}} {{\cal L}^c_n}~,
\label{eq:lsr}
\eeq
 where $\tau$ is the LSR variable, $n$ is the degree of moments, $t_c$ is  the threshold of the ``QCD continuum" which parametrizes, from the discontinuity of the Feynman diagrams, the spectral function  ${\rm Im}\psi_5(t,m_Q^2,\mu)$   where  $\psi_5(t,m_Q^2,\mu)$ is the  (pseudo)scalar correlator:
 \beq
\hspace*{-0.25cm} \psi_{5}(q^2)=i \hspace*{-0.15cm}\int \hspace*{-0.15cm}d^4x ~e^{-iqx}\la 0\vert {\cal T} J_{5}(x)\ga J_{5}(0)\dr^\dagger \vert 0\ra.
 \label{eq:2-pseudo}
 \eeq
\section{QCD expression of the two-point function}
Using the SVZ\,\cite{SVZa} Operator Product Expansion (OPE), the two-point correlator can be written in the form:
\bea
\hspace*{-0.5cm}\psi_5(q^2)&=&\int_{(m_c+m_b)^2}^{\infty}\hspace*{-0.cm}{dt\over t-q^2}\frac{1}{\pi} \mbox{Im}~\psi_5(t,\mu)\vert_{PT} \nnb\\
&+&\la \alpha_s G^2\ra C_{G^2}(q^2,\mu) +\la g^3G^3\ra C_{G^3}(q^2,\mu)+\cdots,
\eea
where $\mu$ is the subtraction scale; $\mbox{Im}~\psi_5(t,\mu)\vert_{PT} $ is the perturbative part of the spectral function; $C_{G^2}$ and $C_{G^2}$ are (perturbatively) calculable Wilson coefficients; $ \la \alpha_s G^2\ra $ and  $\la g^3G^3\ra $ are the non-pertubative gluon condensate of dimensions $d=4,6$ contributions where : $G^2\equiv G^a_{\mu\nu}G^{\mu\nu}_a$ and $g^3G^3\equiv g^3 f_{abc}G_{\mu\nu}^aG^{\nu\rho,b}G_{\rho}^{\mu,c}$.  
As explicitly shown in Ref.\,\cite{BAGAN}, $C_{G^2}$ and $C_{G^2}$  include  the ones of the quark and mixed quark-gluon condensate\,through the relation\,\cite{SVZa,GENERALIS1,BAGAN1}:
\bea
\la \bar QQ\ra&=&-{1\over 12\pi m_Q}\la\alpha_s G^2\ra-{\la g^3 G^3\ra\over 1440\pi^2 m_Q^3}~,\nnb\\
\la \bar QGQ\ra&=&{m_Q\over \pi}\ga\log{m_Q\over \mu}\dr\la\alpha_s G^2\ra-{\la g^3 G^3\ra\over 48\pi^2 m_Q}~,
\eea
from  the heavy quark mass expansion.
\subsection*{\b $q^2=0$ behaviour of the correlator}
 To NLO, the perturbative part of $\psi_{5}(0)$ reads\,\cite{SNB1,SNB2,BECCHI,GENERALIS}:
 \beq
 \psi_{5}(0)\vert_{PT}={3\over 4\pi^2}(m_b+m_c)\ga m_b^3Z_b+m_c^3Z_c\dr~,
 \eeq
 with : 
 \beq
 Z_i=\ga 1- \log{m_i^2\over \mu^2}\dr\ga1+{10\over 3}a_s\dr+{2}a_s \log^2{m_i^2\over \mu^2}~,
 \eeq
 where $i\equiv c,b$;  $\mu$ is the QCD subtraction constant and $a_s\equiv \alpha_s/\pi$ is the QCD coupling. This PT contribution which is present here has to be added to the well-known non-perturbative contribution: 
 \beq
  \psi_{5}(0)\vert_{NP}=-(m_b+m_c)\la \bar cc+\bar bb\ra~,
  \eeq
  for absorbing $log^n(-m_i^2/q^2)$ mass singularities appearing during the  evaluation of the PT two-point function, a technical point not often carefully discussed in some papers.
Working with $\psi_{5}(q^2)$ is safe as $\psi_{5}(0)$, which disappears after successive derivatives, does not affect the pseudoscalar sum rule. This is not the case of the  longitudinal part of the axial-vector two-point function $\Pi^{(0)}_{A}(q^2)$ built from the axial-vector current:
\beq
J_A^\mu\equiv  \bar c(\gamma^\mu\gamma_5) b~,
\eeq
which is
related to $\psi_{5}(q^2)$ through the Ward identity\,\cite{SNB1,SNB2,BECCHI}:
 \beq
 \Pi^{(0)}_{A}(q^2)={1\over q^2}\big{[} \psi_{5}(q^2)-\psi_{5}(0)\big{]}~,
 \eeq
 and  which is also often (uncorrectly) used  in literature.  

\subsection*{\b LO and NLO Perturbative contribution at large $q^2$}
The complete expressions of the PT spectral function has been obtained to LO in\,\cite{FNR}, to NLO in\,\cite{GENERALIS} and explicitly written in \cite{BAGAN}.
It reads ($i\equiv c,b$) :
\bea
&&\hspace*{-0.5cm}\mbox{Im} \psi_5(t)\vert_{PT} =
{3(m_b+m_c)^2\over8\pi t} \bar q{}^4 v \Bigg\{
1+{4\over3}\as \Bigg\{ {3\over8}(7-v^2)\nn\\
&+&\sum_{i=b,c}\Big[ (v+v^{-1})\left(
\dl(\al_1\al_2)-\dl(-\al_i)-\log \al_1 \;\log \be_i
\right)\nnb
\\
&+&
A_i \log \al_i +B_i \log \be_i \Big]\Bigg\}+O(\al_s^2)
\Bigg\}
\eea
%
where
\beq
\dl(x)=-\int_0^x {dy\over y} \log(1-y)
\eeq
and
\bea
A_i&=&{3\over 4}{3m_i+m_j\over m_i+m_j}-{19+2v^2+3v^4\over 32v}\nnb\\
&-&
{m_i(m_i-m_j)\over
\bar q{}^2v(1+v)}\left(1+v+{2v\over 1+\al_i}\right);\nn\\
B_i&=&2+2{m_i^2-m_j^2\over \bar q{}^2 v};
\label{N2}
\\
\al_i&=&{m_i\over m_j}{1-v\over
1+v};\qquad \be_i=\sqrt{1+\al_i}\;{(1+v)^2
\over 4v}\nn\\
\bar q{}^2&=&t-(m_b-m_c)^2;\qquad v=\sqrt{1-4{m_b m_c\over
\bar q{}^2}},
\nn
\eea
where $m_i$ is the on-shell/pole mass. 
\subsection*{\b Higer Orders Perturbative contributions at large $q^2$}
In the absence of a complete result to order $\alpha_s^2$, we shall approximatively use the expression of the spectral function for $m_c=0$:
\beq
\mbox{Im}\psi_5(t)\vert^{N2LO}_{PT}\simeq\frac{1}{8\pi^2}
\as^2 R_{2s}~,
\eeq
where $R_{2s}$ has been obtained semi -analytically in \cite{CHETa,CHETb} and is available as  a Mathematica package program Rvs.m. 

We expect that it is a good approximation because we shall see that the NLO contributions induce (as expected) small corrections in the ratio of moments used to determine $\overline m_{c,b}$ due to the partial cancellation of this contribution . 
 
 We estimate the accuracy of this approximation by comparing this N2LO approximation with the one obtained assuming a geometric growth of the PT coefficients  \cite{SZ}.
  
 We estimate the error due to the truncation of the PT series from the N3LO
contribution estimated, as above, from a geometric growth of the PT series which is expected to mimic the phenomenological $1/q^2$ dimension-two contribution parametrizing the uncalculated large order terms of PT series \cite{CNZa,CNZb,ZAKa,ZAKb}.  

\subsection*{\b $\la\alpha_s G^2\ra$ contribution  at large $q^2$}
We shall use the complete expression of the gluon condensate $\la\alpha_s G^2\ra$ contribution to the two-point correlator given in\,\cite{BAGAN}, which agrees with known results for $m_b=m_c$\,\cite{SNB1,SNB2}. The Wilson cefficient reads: 
\begin{eqnarray}
&&\hspace*{-1.cm}C_{G^2}(q^2,\mu)={1\over\pi}\int_{(m_b+m_c)^2}^\infty
\!\Bigg\{\!\!
-{  m_b m_c \,t\,\left( t\!-\!m_b^2\!-\!m_bm_c\!-\!m_c^2\right)
\over2\,(Q^2\!+\!t)\;[t\!-\!(m_b\!-\!m_c)^2]^{3/2}}\nnb\\
&&\hspace*{-1cm}-{ \sqrt{m_b m_c}}
\Bigg\{
{(m_b\!+\!m_c)^2[t\!-\!(m_b\!+\!m_c)^2]
\over16[Q^2+(m_b+m_c)^2]^2}-
{1\over16[Q^2\!+\!(m_b\!+\!m_c)^2]}
\nonumber\\
&&\hspace*{-1cm}\times \Big[
(m_b\!+\!m_c)^2+{
(
5m_b^2\!+\!18m_b m_c\!+\!5m_c^2
)[t-(m_b+m_c)^2]\over8 m_b m_c}
\Big]
\Bigg\}
\Bigg\}\nonumber\\
&&\hspace*{-1cm}\times{dt\over[t\!-\!(m_b\!+\!m_c)^2]^{5/2}}.
\end{eqnarray}
where $Q^2\equiv -q^2$, 
from which we can easily deduce the Laplace transform.
\subsection*{\b $\la g^3G^3\ra$ contribution at large $q^2$}
 A similar (but lengthy) expression of the $\la g^3G^3\ra$ condensate contribution can also be obtained from \,\cite{BAGAN}, where it has been checked that it agrees with known result for $m_b=m_c$\,\cite{BAGAN1}. It reads :
 \def\S{\Sigma}
\def\SS{{\S^2}}
\begin{eqnarray}
&&\hspace*{-1.5cm}C_{G^3}(q^2)={1\over\pi}\int_\SS^\infty {dt\over[t-\SS]^{9/2}}
\Bigg\{  { m_b m_c\; t\over 6\;
(t+Q^2)\,[t-(m_b-m_c)^2]^{7/2}}
                                                       \nonumber\\
&&\hspace*{-1.5cm}\times
\Big\{   3t^4
 -2(3m_b^2+2m_bm_c+3m_c^2)t^3  
 \nonumber\\
&&\hspace*{-1.5cm}
+  (5m_b^3m_c+18m_b^2m_c^2+5m_bm_c^3)\;t^2\nonumber\\
&&\hspace*{-1.5cm}+   2(3m_b^6+m_b^5m_c-6m_b^4m_c^2-6m_b^3m_c^3-6m_b^2m_c^4+
m_bm_c^5+3m_c^6)\;t     \nonumber\\
&&\hspace*{-1.5cm}-   3(m_b^8+m_b^7m_c-m_b^5m_c^3-2m_b^4m_c^4-m_b^3m_c^5+m_b
m_c^7+m_c^8)
     \Big\} \nonumber\\
&&\hspace*{-1.5cm}- \sqrt{m_bm_c}
\left\{
{-7\S^4\,(t-\SS)^3\over192\,(Q^2+\SS)^4}
\right.\label{ZZZ1}
\nnb\\
&&\hspace*{-1.5cm}+
\left[{7\SS\,(t-\SS)^2\over192}+{A\,(t-\SS)^3\over 1536m_b m_c}\right]
{\SS\over (Q^2+\SS)^3}\nonumber\\
&&\hspace*{-1.5cm}+\left[{-7\S^4\,(t-\SS)\over192}
-{\SS\,A\,(t-\SS)^2\over 1536m_b m_c}
+{B\,(t-\SS)^3\over24576m_b^2m_c^2}\right]{1\over(Q^2+\SS)^2}\nonumber\\
&&\hspace*{-1.5cm}+\left.\left[{7\S^4\over192}+{\SS\,A\,(t-\SS)\over1536m_bm_c}
-{B\,(t-\SS)^2\over24576m_b^2m_c^2}-
{C\,(t-\SS)^3\over196608m_b^3m_c^3} \right] {1\over Q^2+\SS}
\right\}
\Bigg\},\nonumber
\end{eqnarray}
%
with :
\begin{eqnarray}
&&\hspace*{-1.5cm}\Sigma=m_b+m_c\nonumber\\
&&\hspace*{-1.5cm}A=51m_b^2+166m_b m_c+51m_c^2\nnb\\
&&\hspace*{-1.5cm}B=31m_b^4-836m_b^3m_c-1862m_b^2m_c^2-836m_bm_c^3+31m_c^4
\nonumber\\
&&\hspace*{-1.5cm}C=277m_b^4+596m_b^3m_c-514m_b^2m_c^2+596m_bm_c^3+277m_c^4.
\label{ZZZ2}
\end{eqnarray}


\subsection*{\b From On-shell to the $\overline{MS}$-scheme}
We transform the pole masses $m_Q$ to the running masses $\overline m_Q(\mu)$ using the known relation  in the
$\overline{MS}$-scheme to order $\alpha_s^2$ \cite{TAR,COQUEa,COQUEb,SNPOLEa,SNPOLEb,BROAD2a,BROAD2b,CHET2a,CHET2b}:
\bea
m_Q &=& \overline{m}_Q(\mu)\Big{[}
1+{4\over 3} a_s+ (16.2163 -1.0414 n_l)a_s^2\nnb\\
&&+\ln{\mu^2\over m_Q^2} \ga a_s+(8.8472 -0.3611 n_l) a_s^2\dr\nnb\\
&&+\ln^2{\mu^2\over m_Q^2} \ga 1.7917 -0.0833 n_l\dr a_s^2...\Big{]},
\label{eq:pole}
\eea
for $n_l=3$ light flavours. In the following, we shall use $n_f=5$ total number of flavours for the numerical value of $\alpha_s$. 
\section{QCD input parameters}
\nin
The QCD parameters which shall appear in the following analysis will be the charm and bottom quark masses $m_{c,b}$,
 the gluon condensates $ \la\alpha_sG^2\ra$ and $ \la g^3G^3\ra$.  
{\scriptsize
\begin{table}[hbt]
 \caption{QCD input parameters from recent QSSR analysis based on stability critera. The values of $\overline{m}_c(m_c)$ and $\overline{m}_b(m_b)$ come from recent moments SR and their ratios\,\cite{SNmom18} where the errors have been multiplied by a factor 2 to be conservative.}  
\setlength{\tabcolsep}{0.1pc}
    {\small
  \begin{tabular}{llll}
&\\
\hline
\hline
Parameters&Values&Sources& Ref.    \\
\hline
$\alpha_s(M_Z)$& $0.1181(16)(3)$&$M_{\chi_{0c,b}-M_{\eta_{c,b}}}$&Ratios of LSR \, \cite{SNparam}\\
$\overline{m}_c(m_c)$&$1264(12)$ MeV &${J/\psi}$ family&Mom.\,\cite{SNmom18}\\
$\overline{m}_b(m_b)$&$4188(16)$ MeV&${\Upsilon}$  family&Mom.\,\cite{SNmom18}\\
$\la\alpha_s G^2\ra$& $(6.35\pm 0.35)\times 10^{-2}$ GeV$^4$&Hadrons&QSSR average\,\cite{SNparam}\\
$\la g^3  G^3\ra$& $(8.2\pm 2.0)$ GeV$^2\times\la\alpha_s G^2\ra$&$J/\psi$  family&Mom. \cite{SNH10a,SNH10b}\\
&&&Ratios of LSR \cite{SNH10c}\\
\hline\hline
\end{tabular}
}
\label{tab:param}
\end{table}
} 
\subsection*{\b QCD coupling $\alpha_s$}
We shall use from the $M_{\chi_{0c}}-M_{\eta_{c}}$ mass-splitting sum rule\,\cite{SNparam}: 
 \bea&&\hspace*{-1cm} 
 \alpha_s(2.85)=0.262(9) \leadsto\alpha_s(M_\tau)=0.318(15)\nnb\\
& \leadsto&\alpha_s(M_Z)=0.1183(19)(3)
\eea
which is more precise than the one from  $M_{\chi_{0b}}-M_{\eta_{b}}$\,\cite{SNparam} : 
\bea &&\hspace*{-1cm} 
 \alpha_s(9.50)=0.180(8) \leadsto\alpha_s(M_\tau)=0.312(27)\nnb\\
&&\leadsto\alpha_s(M_Z)=0.1175(32)(3).
 \eea
 These lead to the mean value quoted in Table\,\ref{tab:param}, which is
 in complete agreement with the world average\,\cite{PDG}:
\beq
\alpha_s(M_Z)=0.1181(11),
\eeq
but with a larger error. 
\subsection*{\b $c$ and $b$ quark masses}
For the $c$ and $b$ quarks, we shall use the recent determinations\,\cite{SNmom18} of  the running masses and the corresponding value of $\alpha_s$ evaluated at the scale $\mu$ obtained using the same sum rule approach from charmonium and bottomium systems. The range of values given in Table \ref{tab:param} enlarged by a factor 2 are within  the PDG average\,\cite{PDG}. 
\subsection*{\b Gluon and quark-gluon mixed condensates}
For the $\la \alpha_s G^2\ra$ condensate, we use the recent estimate obtained from a correlation with the values of the heavy quark masses and $\alpha_s$ which can be compared with the QSSR average from different channels\,\cite{SNparam}.  

The one of $ \la g^3G^3\ra$ comes from a QSSR analysis of charmonium systems. Their values are given in Table\,\ref{tab:param}. 

\section{Parametrisation of the spectral function}
-- In the present case, where no complete data on the $B_c$ spectral function are available, we use the duality ansatz:
 \beq
 {\rm Im} \psi_5(t)\simeq f_P^2 M_P^4 \delta(t-M_P^2) +\Theta(t-t_c) ``QCD~{\rm  continuum}",
 \eeq
 for parametrizing the spectral function. $M_P$ and $f_P$ are the lowest ground state mass and coupling analogue to $f_\pi$. The ``QCD continuum" is the imaginary part of the QCD two-point function while $t_c$ is its threshold. Within a such parametrization, one  obtains: 
 \beq
  {\cal R}^{c}_n\equiv {\cal R}\simeq M_P^2~,
  \label{eq:mass}
  \eeq
 indicating that the ratio of moments appears to be a useful tool for extracting the mass of the hadron ground state\,\cite{SNB1,SNB2,SNB3,SNB4,SNREV15}. 
 
 -- This simple model has been tested in different channels where complete data are available (charmonium, bottomium and $e^+e^-\to I=1$ hadrons)\,\cite{SNB1,SNB2,BERTa}. It was shown that, within the model, the sum rule  reproduces well the one using the complete data, while
the masses of the lowest ground state mesons ($J/\psi,~\Upsilon$ and $\rho$) have been predicted with a good accuracy.  In the extreme case of the Goldstone pion, the sum rule using the spectral function parametrized by this simple model\,\cite{SNB1,SNB2} and the more complete one by ChPT\,\cite{BIJNENS} lead to  similar values of the sum of light quark masses $(m_u+m_d)$ indicating the efficiency of this simple parametrization. 

-- An eventual violation of the quark-hadron duality (DV)\,\cite{SHIF,PERIS} has been frequently tested  in the accurate determination of $\alpha_s(\tau)$ from hadronic $\tau$-decay data\,\cite{PERIS,SNTAU,PICH}, where its quantitative effect in the spectral function was found to be less than 1\%. Typically, the DV has the form: 
\beq
\Delta{\rm Im} \psi_5(t)\sim (m_c+m_b)^2t~{\rm e}^{-\kappa t} {\rm sin} (\alpha+\beta t) \theta (t-t_c)~,
\eeq 
where $\kappa,\alpha,\beta$ are model-dependent  fitted parameters but not based from first principles. Within this model, where the contribution is doubly exponential suppressed in the Laplace sum rule analysis, we expect that in the stability regions where the QCD continuum contribution to the sum rule is minimal and where the optimal results in this paper will be extracted, such duality violations can be safely neglected. 

-- Therefore, we (a priori) expect that one can extract with a good accuracy the $c$ and $b$ running quark masses and  the $B_c$ decay constant within the approach. An eventual improvement of the results can be done after a more complete measurement of the $B_c$ pseudoscalar spectral function which is not an easy task though the recent discovery by CMS\,\cite{CMS} of the $B_c(2S)$ state at 6872(1.5)\,MeV is a good starting point in  this direction. 
 
-- In the following, in order to minimize the effects of unkown higher radial excitations smeared by the QCD continuum and some eventual quark-duality violations, we shall work with the lowest ratio of moments $ {\cal R}^{c}_0$ for extracting the quark masses and with the lowest moment $ {\cal L}^c_0 $ for estimating the decay constant $f_{B_c}$. Moment with negative $n$ will not be considered due to their sensitivity on the non-perturbative contributions such as $\psi_5(0)$. 
  
\section{Optimization Criteria}
For extracting the optimal results from the analysis, we shall use optimization criteria (minimum sensitivity) of the observables versus the variation  of the external variables namely the $\tau$ sum rule parameter, the QCD continuum threshold $t_c$ and the subtraction point $\mu$. 

Results based on these criteria have lead to successful predictions in the current literature\,\cite{SNB1,SNB2}. $\tau$-stability has been introduced and tested by Bell-Bertlmann using the toy model of harmonic oscillator\,\cite{BERTa} and applied successfully in the heavy\, \cite{BELLa,BELLb,BERTa,BERTb,BERTc,BERTd,NEUF,SHAW,SNcb3,SNHeavy,SNHeavy2,SNHQET13}  and light quarks systems\,\cite{SVZa,SVZb,SNB1,SNB2,SNB3,SNB4,SNREV15,SNL14}. It has been extended later on to the $t_c$-stability\,\cite{SNB1,SNB2,SNB3,SNB4} and  to the $\mu$-stability criteria\,\cite{SNp13,SNHQET13,SNL14,SNp15,SNparam}. 

Stability on the number $n$ of heavy quark moments have also been used\,\cite{SNH10a,SNH10b,SNH10c,SNmom18}. 

One can notice in the previous works that these criteria have lead to more solid theoretical basis and noticeable improved results from the sum rule analysis. 

\section{
$\overline{m}_c$ and $\overline{m}_b$ 
 from $M_{B_c}$}
\nin
In the following, we look for the stability regions of the external parameters $\tau, t_c$ and $\mu$ where we shall extract our optimal result.
\subsection*{\b $\tau$ stability}

\begin{figure}[hbt]
\vspace*{-0.25cm}
\begin{center}
\includegraphics[width=11cm]{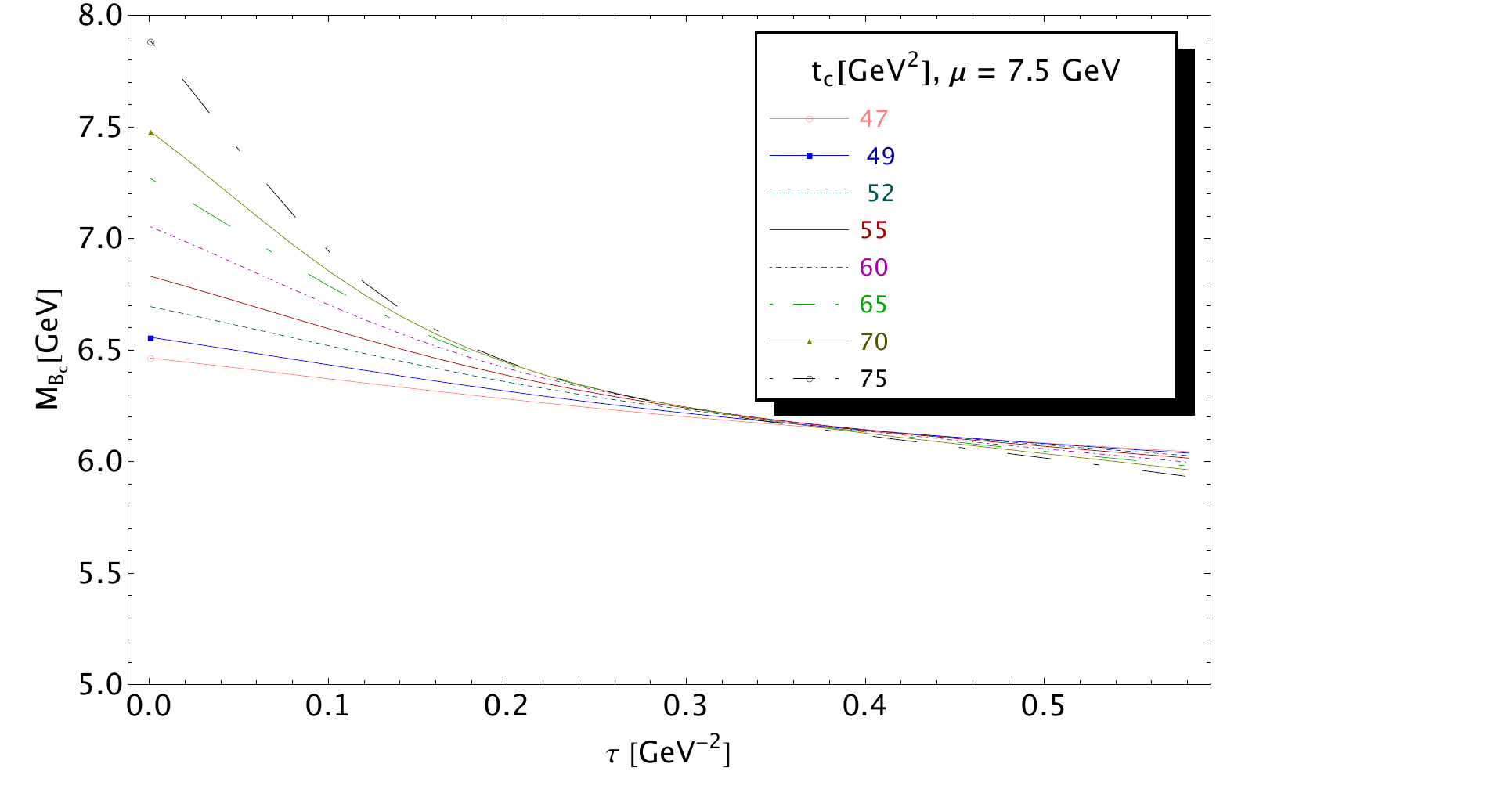}
\vspace*{-0.5cm}
\caption{\footnotesize  $M_{B_c}$ as function of $\tau$ for different values of $t_c$, for $\mu$=7.5 GeV and for given values of $\overline m_{c,b}(\overline m_{c,b})$ in Table\,\ref{tab:param}.} 
\label{fig:bctau}
\end{center}
\end{figure} 
In a first step, fixing  the value of $\mu=7.5$ GeV which we shall justify later and which is the central value of  $\mu=(7.5\pm 0.5)$ GeV obtained in\,\cite{SNp15}, we show in Fig.\,\ref{fig:bctau} the behaviour of $M_{B_c}$ for different values of $t_c$ where the central values of $\overline{m}_c(\overline{m}_c)$=1264 MeV  and $\overline{m}_b(\overline{m}_b)$=4188 MeV given in Table\,\ref{tab:param} have been used.
We see that the inflexion points  at $\tau\simeq (0.30\sim 0.32)$ GeV$^{-2}$ appear for $t_c\geq$ 52 GeV$^2$. 
The smallest value of $\sqrt{t_c}$ is around the $B_c(2S)$ mass of 6872(1.5) MeV recently discovered by CMS\,\cite{CMS} but does not necessarily co\"\i ncide with it as the QCD continuum is expected to smear all higher states contributions to the spectral function. Instead, its value is related by duality  to the ground state parameters as discussed in\,\cite{LAUNERb} from a FESR analysis of the $\rho$-meson channel. 
\subsection*{\b $t_c$ stability}
\begin{figure}[hbt]
\vspace*{-0.5cm}
\begin{center}
\includegraphics[width=9cm]{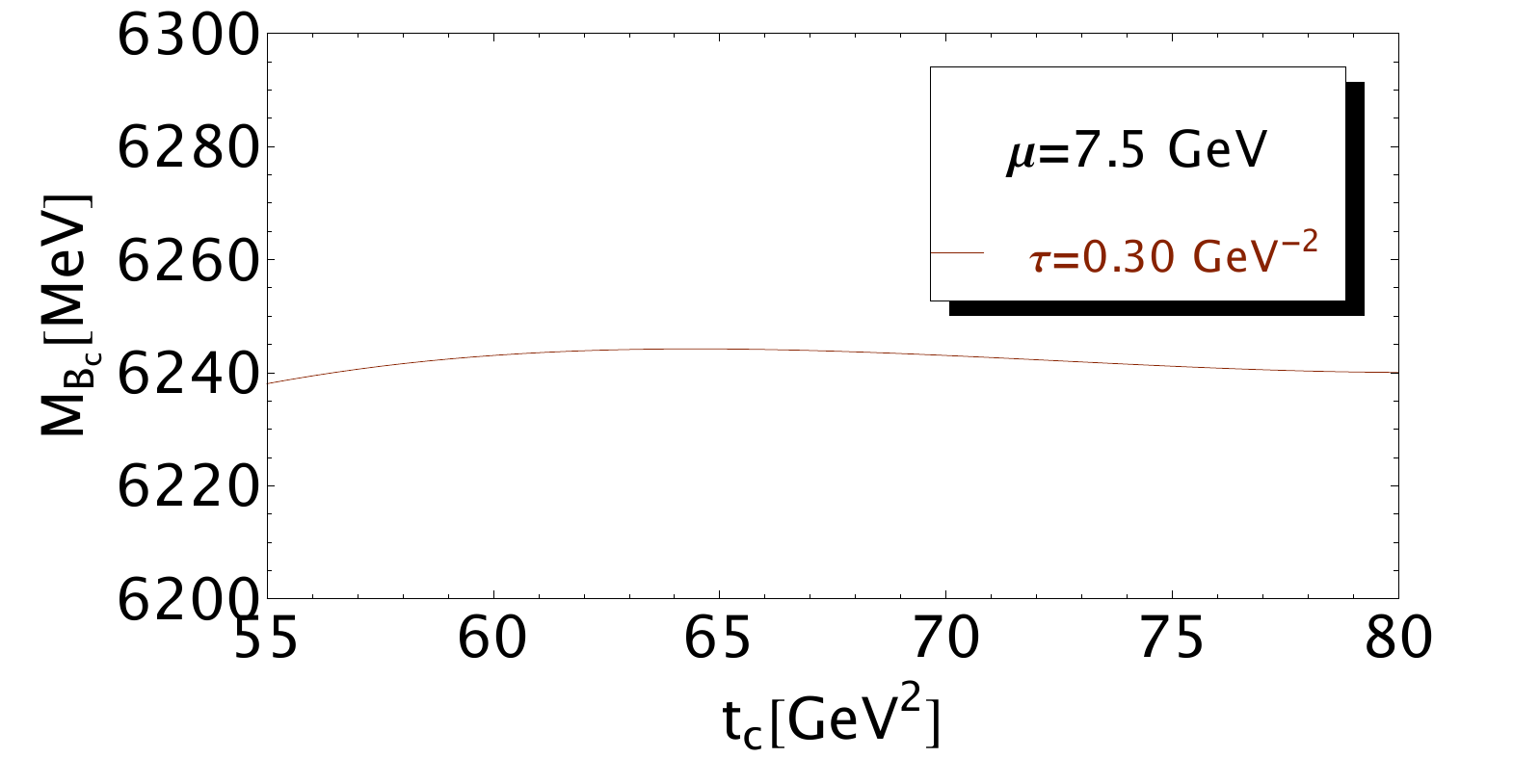}
\vspace*{-0.5cm}
\caption{\footnotesize  $M_{B_c}$ as function of $t_c$ for $\mu$=7.5 GeV and for the range  of $\tau$-stability values. We use the central values of $\overline m_{c,b}(\overline m_{c,b})$ given in Table\,\ref{tab:param}.} 
\label{fig:mbc-tc}
\end{center}
\end{figure}
\begin{figure}[hbt]
\vspace*{-0.25cm}
\begin{center}
\includegraphics[width=10cm]{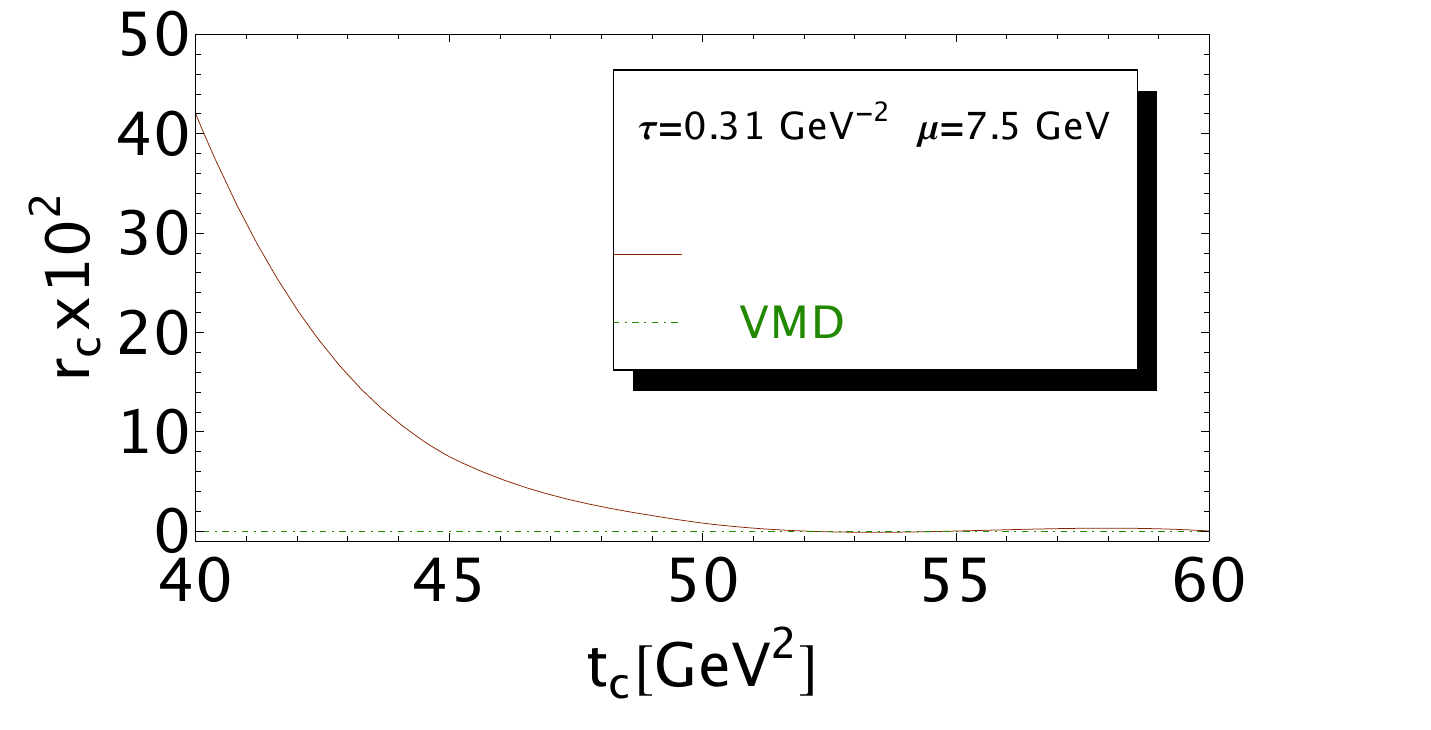}
\vspace*{-0.5cm}
\caption{\footnotesize  Ratio $r_c$ of the continuum over the lowest ground state contribution as function of $t_c$ at the corresponding $\tau$-stability points for $ \mu$=7.5 GeV and for given values of $\overline m_{c,b}(\overline m_{c,b})$ in Table\,\ref{tab:param}. The dashed-dotted line is the contribution for a Vector Meson Dominance (VDM) assumption of the spectral function. } 
\label{fig:cont}
\end{center}
\end{figure}

 
We show in Fig.\,\ref{fig:mbc-tc} the behaviour of $M_{B_c}$ versus $t_c$ which is very stable. For definiteness, we take $t_c$ in the range 52 to 79 GeV$^2$ where we have a slight maximum at $t_c\simeq 60$ GeV$^2$. At this range of $t_c$ values, one can easily check that the QCD continuum contribution to the sum rule is (almost) negligible. 
To have more insights on this contribution, we show in Fig.\,\ref{fig:cont} the ratio of the continuum over the lowest ground state contribution as predicted by QCD :
\beq
r_c\equiv{\int_{t_c}^\infty dt{\rm e}^{-t\tau}\Delta{\rm Im} \psi_5^{cont}\over\int_{(m_c+m_b)^2}^{t_c}dt{\rm e}^{-t\tau}\Delta{\rm Im} \psi_5^{B_c}}
\eeq
where one can indeed see that the QCD continuum to the moment sum rule ${\cal L}_0^c$ is negligible in this range of $t_c$ values. This contribution is 
is even less in the ratio of moments ${\cal R}_0^c$ used to get $M_{B_c}$.  
\subsection*{\b $\mu$ stability}
\begin{figure}[hbt]
\vspace*{-0.25cm}
\begin{center}
\includegraphics[width=9cm]{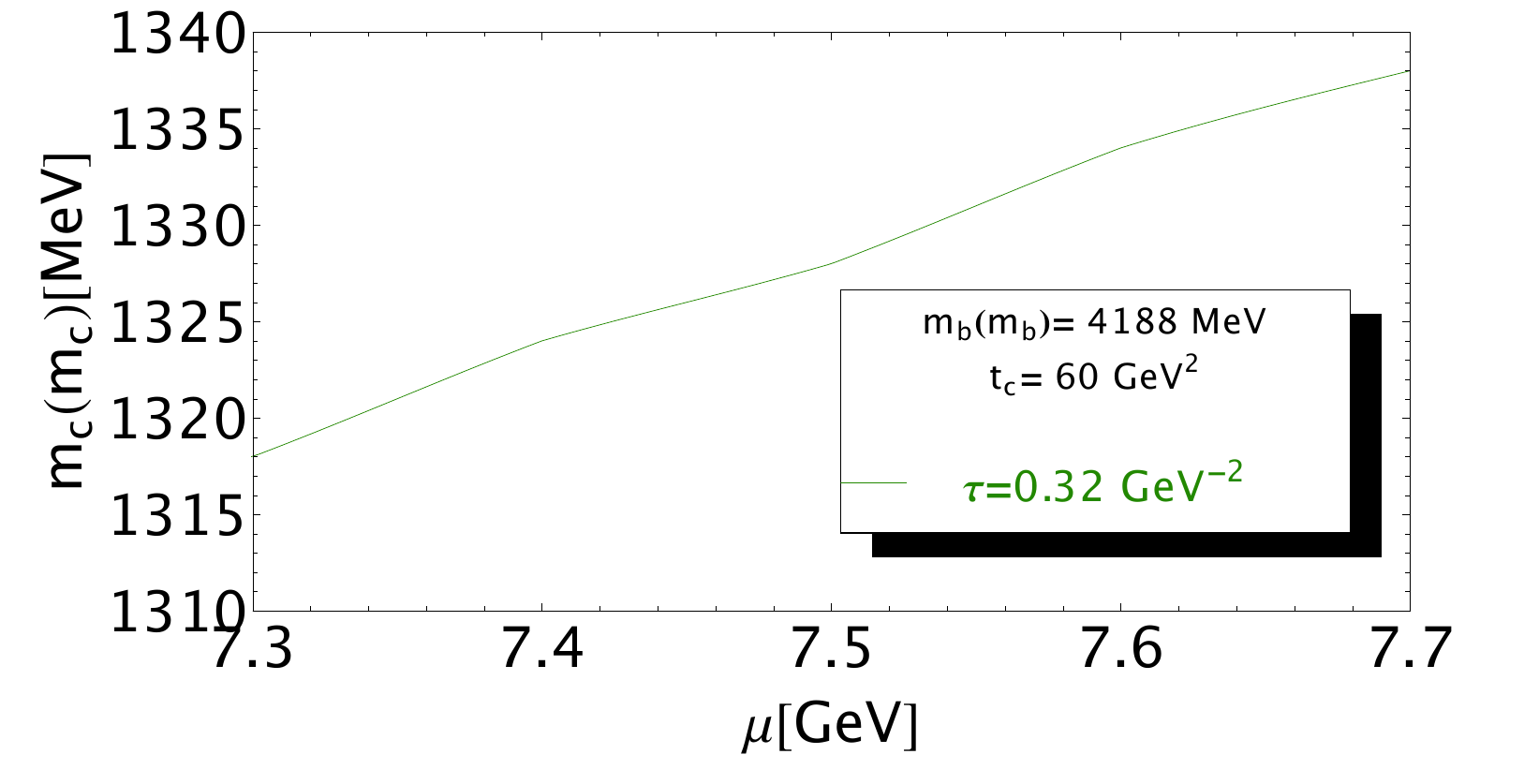}
\vspace*{-0.5cm}
\caption{\footnotesize  $\overline m_{c}(\overline m_{c})$ as function of $\mu$ for $\tau\simeq 0.32$ GeV$^2$ and for the central value of $\overline m_{b}(\overline m_{b})$ given in Table\,\ref{tab:param}.} 
\label{fig:mc-mu}
\end{center}
\end{figure}

Given e.g the central value of $\overline{m}_b(\overline{m}_b)=4188$ MeV from Table\,\ref{tab:param} and using $\tau=.32$ GeV$^{-2}$ and $t_c=60$ GeV$^2$, we show in Fig.\,\ref{fig:mc-mu} the correlated values of $\overline{m}_c(\overline{m}_c)$ at different values of $\mu$ needed for reproducing $M_{B_c}$. We obtain an inflexion point at :
\beq
\mu = (7.5\pm 0.1) ~{\rm GeV}~,
\label{eq:mu}
\eeq
which we shall use  in the following. This value agrees with the one $\mu=(7.5\pm 0.5)$ GeV quoted in\,\cite{SNp15} using different ways. 

\subsection*{\b Extracting the set $(\overline{m}_c,\overline{m}_b) $}
In the following, we study the correlation between $\overline{m}_c$ and $\overline{m}_b$ needed for reproducing the experimental mass\,\cite{PDG} :
\beq
M^{exp}_{B_c}=6275.6(1.1)~{\rm MeV}~,
\eeq 
from the ratio ${\cal R}^c_0$ of Laplace sum rules defined in Eqs.\,\ref{eq:lsr} and \ref{eq:mass}.

\begin{figure}[hbt]
\vspace*{-0.25cm}
\begin{center}
\includegraphics[width=9cm]{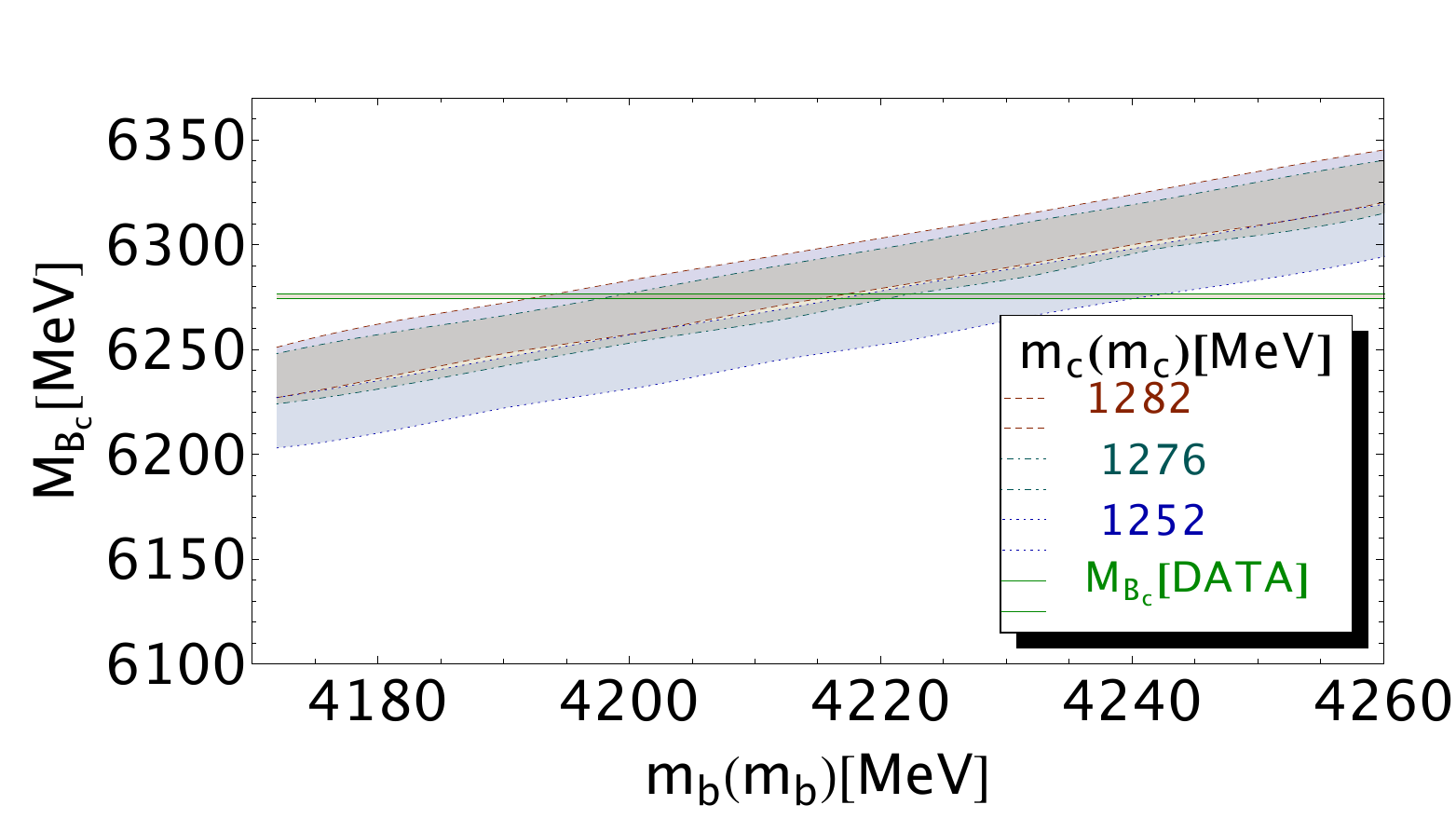}
\vspace*{-0.5cm}
\caption{\footnotesize $M_{B_c}$ as function of $\overline m_b(\overline m_b)$ for different values of $\overline m_c(\overline m_c)$,  for $\mu$=7.5 GeV and for the range  of $\tau$-stability values $\tau=(0.30-0.32)$ GeV$^{-2}$. }
\label{fig:mc-mb}
\end{center}
\end{figure}

-- Allowing $\overline{m}_c(\overline{m}_c)$ to move in the range :
\beq
\overline{m}_c(\overline{m}_c)\simeq (1252-1282)~{\rm MeV}
\eeq
from the $J/\psi$ and $M_{\chi_{0c}-M_{\eta_c}}$ mass-splitting sum rules,
we show in Fig.\,\ref{fig:mc-mb}, the predictions for $M_{B_c}$ as a function
of $\overline{m}_b(\overline{m}_b)$. The band is the error induced
by the choice of the stability points $\tau=(0.30-0.32)$ GeV$^{-2}$ which is about (12-13) MeV, while the error due to some other parameters are negligible.
Then, we deduce : 
\bea
\overline m_b(\overline m_b)=(4195-4245)~{\rm MeV}~,
\eea
which leads to the correlated set of values in units of MeV:
\beq
[\overline m_b(\overline m_b), \overline{m}_c(\overline{m}_c)]=[4195,1282]~...~[4245,1252]~,
\eeq
This result shows that a small value of $m_c$ is correlated to a large value of $m_b$ and vice-versa.

-- Scrutinizing Fig.\,\ref{fig:mc-mb}, one can see that, at fixed $m_b$, e.g 4220 MeV,  $M_{B_c}$ increases  with the $m_c$ values given in the legend (vertical line), as intuitively expected. On the other, fixing the value of $m_c$  at the one  in the legend say 1282 MeV, on can see (straightline with a positive slope) that $M_{B_c}$ increases when $m_b$ increases on the $m_b$-axis as also intuitively expected. 

-- Considering that the values of $\overline{m}_b(\overline{m}_b)$ are inside the range:
\beq
\overline m_b(\overline m_b)=(4176-4209)~{\rm MeV}~,
\label{eq:mb1}
\eeq 
allowed from the $\Upsilon$ sum rules as given in Table\,\ref{tab:hmass}, we can deduce from Fig.\,\ref{fig:mc-mb} stronger constraints on $\overline{m}_b(\overline{m}_b)$ :
\bea
\overline m_b(\overline m_b)&=&(4195-4209) ~{\rm MeV}\nnb\\
&=&
4202(7)~{\rm MeV}~,
\label{eq:mb}
\eea
where the error is similar to the accurate value from the $\Upsilon$ sum rule in Table\,\ref{tab:hmass}. This is due to the small intersection region of the results from the $J/\psi,~\Upsilon$ and the $B_c$ sum rules. 
With this range of values, we deduce :
\bea
\overline m_c(\overline m_c)&=&1286(8)_{fig.4}(14)_\tau(1)_{t_c}~{\rm MeV}~,\nnb\\
&=& 1286(16)~{\rm MeV}~,
\label{eq:mc}
\eea 
 where the errors due to some other parameters and to the truncation of the PT series are negligible. The susbscript $fig.4$ indicates that the error comes from the intersection region between the $\Upsilon$ and $B_c$ sum rules in Fig.\,\ref{fig:mc-mb}.  
 
We consider the values in Eqs.\,\ref{eq:mb} and \,\ref{eq:mc} as our final determinations of $\overline m_b(\overline m_b)$ and $\overline m_c(\overline m_c)$ from $M_{B_c}$ and combined constraints from the $J/\psi$ and $\Upsilon$ sum rules. 

\subsection*{\b Comments}

-- One can notice that the effect of the PT radiative corrections are quite small in the ratio of moments because the ones of the absolute moments ${\cal L}_{0,1}$  tend to compensate each others. This fact can be checked from a numerical parametrization of the LSR ratio. At the optimization scale $\tau\simeq 0.32 $ GeV$^{-2}$ and $\mu=7.5$ GeV, it reads\,($a_s\equiv \alpha_s/\pi$) :
\beq
\sqrt{\cal R}_0|^{N2LO}_{PT}\simeq \sqrt{\cal R}_0|^{LO}_{PT}(1-0.16a_s-0.42a_s^2)~,
\eeq
while  the LSR lowest moment  is :
\beq
\sqrt{\cal L}_0|^{N2LO}_{PT}\simeq\sqrt{\cal L}_0|^{LO}_{PT}(1+6a_s+26.4a_s^2)~.
\label{eq:n2lo}
\eeq

-- One can also notice that the approximate N2LO contribution obtained for $m_c=0$ in the lowest moment is about the same as the one $36a_s^2$ which one would obtain using a geometric growth of  the $a_s$ PT coefficients\,\cite{CNZb}. Therefore, the error induced by the difference of the two N2LO approximations is negligble. 

-- The contribution of the gluon condensates is also small at the optimization scale as $\la \alpha_s G^2\ra$ increases $M_{B_c}$ by about 5 MeV while $\la g^3 G^3\ra$ decreases it by 1 MeV. These contributions are small and also  show the good convergence of the OPE. 
Then, its induces an increase of about 6 MeV in the quark mass values and introduces a negligigle error of 1 MeV.   However, the non-perturbative contributions are important for having the $\tau$-stability region. 

-- As the QCD continuum contribution which is expected to smear all radial excitation contributions is negligible at the optimization region due to the exponential dumping factor of the sum rule, we expect that some eventual DV discussed previously can be safely neglected due to its  doubly exponential suppression in the LSR analysis. We also expect that  the effects of higher radial excitations can be similarly neglected like the one of the QCD continuum .
\section{Comparison with some other QSSR determinations }
We compare the previous results with the ones in Table\,\ref{tab:hmass} obtained from some other QSSR analysis using the same stability 
criteria.  A tentative average of the central values and using the error from the most precise predictions from $J/\psi$ and $\Upsilon$ families leads to the averages quoted in Table\,\ref{tab:hmass}.
{\scriptsize
\begin{table}[hbt]
 \caption{Values of $\overline{m}_c(m_c)$ and $\overline{m}_b(m_b)$ coming from our recent QSSR analysis based on stability criteria. Some other determinations can be found in\,\cite{PDG}. }  
\setlength{\tabcolsep}{0.1pc}
    {\small
  \begin{tabular}{llll}
&\\
\hline
\hline
Parameters&Values [MeV]& Sources & Ref.    \\
\hline
$\overline{m}_c(m_c)$&$1256(30)$  &${J/\psi}$ family&Ratios of LSR17\, \cite{SNparam}\\
&$1266(16)$ &$M_{\chi_{0c}-M_{\eta_c}}$&Ratios of LSR\, \cite{SNparam}\\
&1264(6) &${J/\psi}$ family& MOM \& Ratios of MOM\, \cite{SNmom18} \\
&1286(66)&$M_D$ & Ratios of LSR\,\cite{SNp13}\\
&1286(16)&$M_{B_c}$ & Ratios of LSR {\it (This work)} \\
&{\it 1266(6)}&{\it Average}& {\it This work} \\
\\
$\overline{m}_b(m_b)$&$4192(17)$ MeV&${\Upsilon}$ family&Ratios of LSR\,\cite{SNparam}\\
&4188(8) &$\Upsilon$ family &MOM \& Ratios of MOM\, \cite{SNmom18} \\
&4236(69) &$M_B$&Ratios of MOM \& of LSR\,\cite{SNp13}\\
&4213(59)&$M_B$& Ratio of HQET-LSR\,\cite{SNHQET13}\\
&4202(7)&$M_{B_c}$ & Ratios of LSR{\it (This work)}\\
&{\it 4196(8)}&{\it  Average}& {\it This work}\\
\hline\hline
\end{tabular}
}
\label{tab:hmass}
\end{table}
} 

\section{Revisiting   $f_{B_c}$}
\nin
Using the previous correlated values of $(\overline{m}_c,\overline{m}_b)$, we reconsider the estimate of $f_{B_c}$ recently done in Ref.\,\cite{SNp15}. 
\subsection*{\b $\tau$ and $t_c$ stabilities}
\begin{figure}[hbt]
\vspace*{-0.25cm}
\begin{center}
\includegraphics[width=11cm]{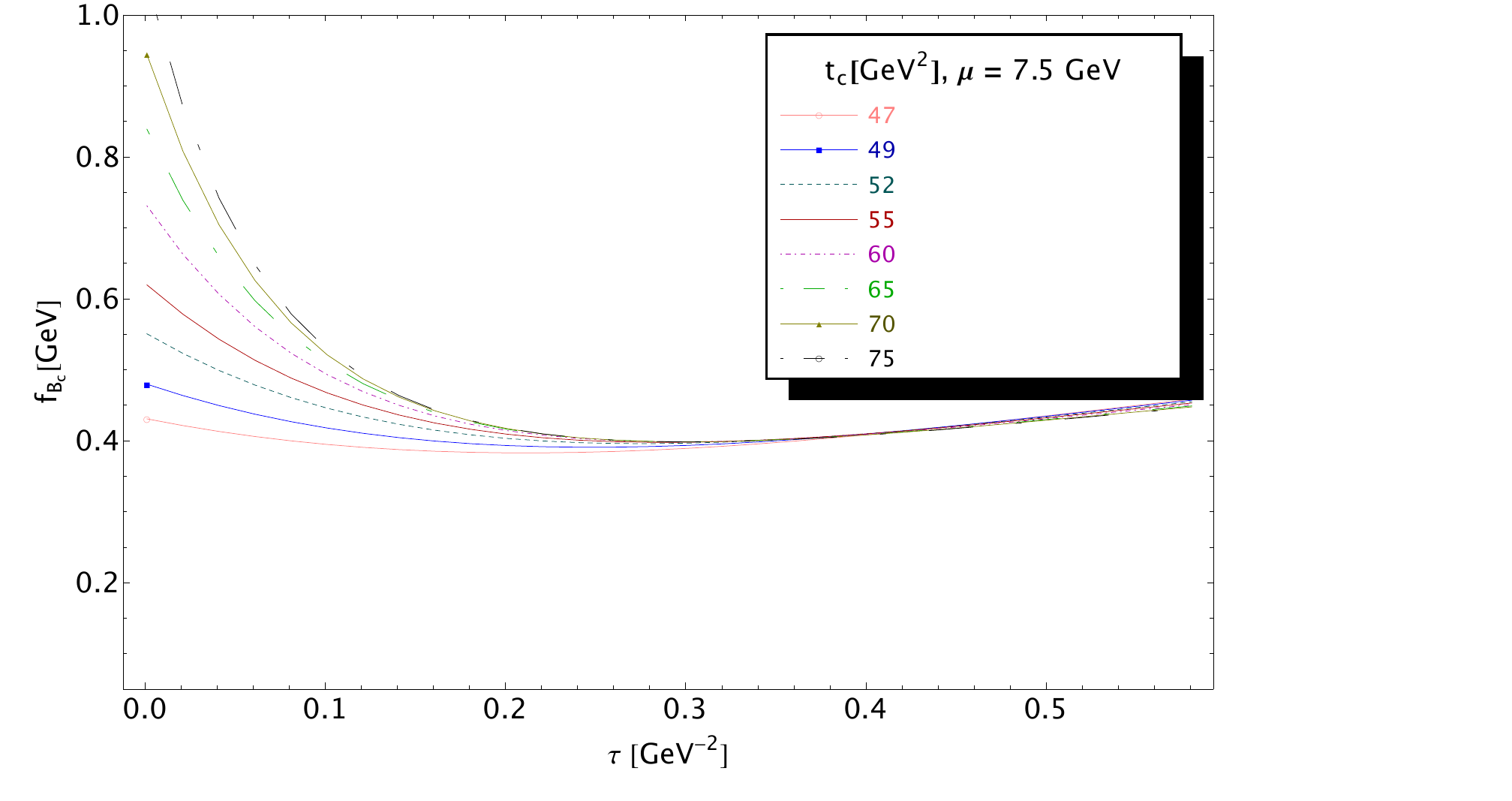}
\vspace*{-0.5cm}
\caption{\footnotesize  $f_{B_c}$  versus $\tau$ given $\mu=7.5$ GeV and $[\overline{m}_c(\overline{m}_c),\overline{m}_b(\overline{m}_b)]=[1264,4188]$ MeV from Table\,\ref{tab:param}.} 
\label{fig:fbc}
\end{center}
\end{figure}

We show the $\tau$-behaviour of $f_{B_c}$ in Fig.\,\ref{fig:fbc} for different values of $t_c$ for $\mu=7.5$ GeV and for $[\overline{m}_c(\overline{m}_c),\overline{m}_b(\overline{m}_b)]=[1264,4188]$ MeV from Table\,\ref{tab:param}. We start to have $\tau$-stability (minimum) for the set [$t_c,\tau$]=[47 GeV$^2$,0.22 GeV$^{-2}$] and $t_c$-stability for the set [$60$ GeV$^2$, $(0.30-0.32)$ GeV$^{-2}$]. One can notice that the $\tau$-stability starts earlier for smaller $t_c$-value than to the case of the ratio of moments used in the preceeding sections.  To be conservative, we shall consider the value  of $f_{B_c}$ obtained inside this larger range of $t_c$-values and take as a final value its mean.  
In this range, the value of $f_{B_c}$ increases by about 15 MeV.
\begin{figure}[hbt]
\vspace*{-0.25cm}
\begin{center}
\includegraphics[width=10cm]{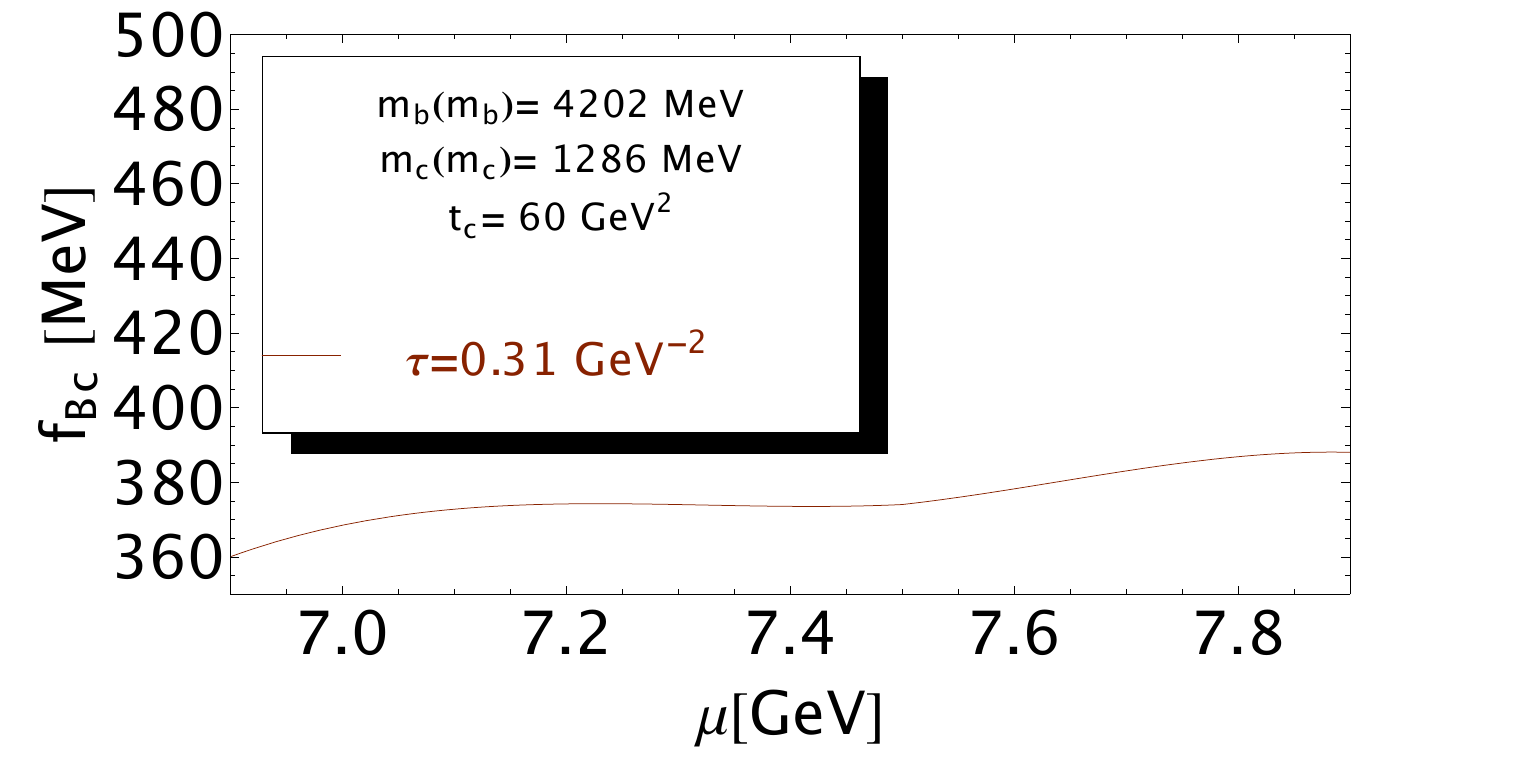}
\vspace*{-0.5cm}
\caption{\footnotesize  $f_{B_c}$ as function of $\mu$ for $\tau\simeq 0.31$ GeV$^{-2}$ and for the central value of $\overline m_{c,b}(\overline m_{c,b})$ given in Eqs.\,\ref{eq:mb} and\,\ref{eq:mc}.} 
\label{fig:fbc-mu}
\end{center}
\end{figure} 
\subsection*{\b $\mu$ stability}
We study the influence of $\mu$ on $f_{B_c}$ in Fig.\,\ref{fig:fbc-mu} given the values of $\tau$ and $m_{b,c}$. We see a clear stability for $\mu=(7.2-7.5)$ GeV which is consistent with the one for $M_{B_c}$ and with the one obtained in \cite{SNp15} indicating the self-consitency of the analysis. 
\subsection*{\b Higher orders (HO) PT corrections}
-- The N2LO contribution increases the prediction from LO $\oplus$ NLO by 24 MeV.
We estimate the error induced by using the result at $m_c=0$ by comparing it with the one obtained from the estimate of the N2LO coefficient using a geometric growth of the PT series (see Eq.\,\ref{eq:n2lo}). The difference induces an error of $\pm 9.6a_s^2$ which corresponds to $\pm 9$ MeV. 

-- We estimate the error due to the uncalculated higher order (HO) part of the PT series from the N3LO contribution estimated by using the geometric growth of the coefficients  given in the numerical expression in Eq.\,\ref{eq:n2lo} which is $\pm 158a_s^3$. It introduces an error of about 10 MeV.

\subsection*{\b Gluon condensate contributions}
The inclusion of the $\la \alpha_s G^2\ra$ condensate increases the sum of the PT contributions by 3 MeV, while the inclusion of the $\la g^3 G^3\ra$ dcreases the prediction by 1 MeV. These contributions and the induced error are negligible. 

\subsection*{\b Result}
-- The result of the analysis in units of MeV is:
\bea
f_{B_c}&=&368(1)_\tau(8)_{t_c}(7)_{m_{c}}(5)_{m_{b}}(1)_{\alpha_s}(1)_\mu 9)_{N2LO}(10)_{HO}\nnb\\
&=& 368(18)~{\rm MeV},
\label{eq:fbc1}
\eea
if one uses the mass values obtained in Eqs.\,\ref{eq:mb} and \ref{eq:mc}, while it is : 
\bea
f_{B_c}&=&381(1)_\tau(8)_{t_c}(3)_{m_{c}}(5)_{m_{b}}(1)_{\alpha_s}(1)_\mu(9)_{N2LO}(10)_{HO}\nnb\\
&=& 381(17)~{\rm MeV},
\label{eq:fbc2}
\eea
if one uses the tentative mass averages given in Table\,\ref{tab:hmass}. We take as a final result the mean of the two determinations:
\beq
f_{B_c}=371(17)~{\rm MeV}~.
\label{eq:fbc}
\eeq
This result improves the previous one $f_{B_c}=436(40)$ MeV obtained recently  in Ref.\,\cite{SNp15} and the earlier results in\,\cite{SNBc,COL2,CHABAB}. It confirms and improves the one $f_{B_c}=388(29)$ MeV averaged from moments and LSR in\,\cite{BAGAN} where the values of the pole masses have been used. However, it disagrees with some results including  the lattice one reviewed in Table\,3 of\,\cite{SNp15}.  New estimates from the lattice approach is needed for clarifying the issue. Comments related to some of the previous works have been already addressed in\,\cite{SNp15} and can be consulted there. 
\section{Attempted upper bound for $f_{B_c(2S)}$}
We attempt to give an upper bound to $f_{B_c(2S)}$ by using a `` two resonances + QCD continuum" parametrization of the spectral function.
However, we are aware on the fact that due to the exponential suppression of the $B_c(2S)$ contribution compared to $B_c(1S)$ and of its eventual smaller coupling as expected for the observed radial excitations in some other channels, we may not extract  with a good precision the 
 $B_c(2S)$ decay constant from this approach. Instead by using the positivity of the QCD continuum contribution for  $t_c\geq 47~ {\rm GeV}^2$ just above $M_{B_c(2S)}^2$, one obtains the semi-analytic expression from ${\cal L}_0^c$ :
 \beq
\rho_{B_c}\equiv{\ga f_{B_c(2S)}\over f_{B_c}\dr^2}{\ga M_{B_c(2S)}\over M_{B_c}\dr^4{\rm e}^{-\ga M_{B_c(2S)}^2-M_{B_c}^2\dr{\tau}}}\leq 3.6\%,
\eeq
at the $\tau$-stability of about $0.22~ \rm{GeV}^{-2}$  as can be deduced from Fig.\,\ref{fig:cont}. Using the previous value of $f_{B_c}$ in Eq.\,\ref{eq:fbc}, we deduce:
 \beq
 f_{B_c(2S)}\leq (139\pm 6)~\rm{MeV}~,
 \label{eq:fbc2s}
 \eeq
 indicating that the radial excitation couples weaker to the corresponding quark current than the ground state meson. This feature has been already observed experimentally in the case of light ($\pi,\rho$) and heavy ($\psi,\Upsilon$) mesons. 
\section{Summary and Conclusions }
\subsection*{\b $\overline{m}_c$ and $\overline{m}_b$}
We have used QCD Laplace sum rules to estimate ({\it for the first time}) the correlated values of $\overline{m}_c(\overline{m}_c)$ and $\overline{m}_b(\overline{m}_b)$ from the $B_c$-meson mass allowing them to move inside the extended (multiplied by a factor 2: see Fig\,\ref{fig:mc-mb} and Table\,\ref{tab:param}) range of values allowed by charmonium and bottomium sum rules. These values :
$$
\overline{m}_b(\overline{m}_b)=4202(7) ~{\rm MeV} ~ ({\rm Eq}.\,\ref{eq:mb}), 
$$
$$
\overline{m}_c(\overline{m}_c) = 1286(16)~{\rm MeV}~ ({\rm Eq}.\,\ref{eq:mc}),
\label{eq:mc-mean}
$$
agree with previous recent ones from charmonium and bottomium systems quoted in Table\,\ref{tab:hmass}. The errors are similar to the ones from $J/\psi$ and $\Upsilon$ sum rules. They have been relatively reduced compared to the ones from the $D$ and$B$ meson masses thanks to the extra constraints on the range of variations of $\overline m_{c,b}(\overline m_{c,b})$ used in Fig.\,\ref{fig:mc-mb} from $J/\psi$ and $\Upsilon$ sum rules. Using these values and the ones from recent different QSSR determinations collected in Table\,\ref{tab:hmass}, we deduce the QSSR average:
$$
\overline{m}_c(\overline{m}_c) = 1266(6)~{\rm MeV} ~{\rm and}~ \overline{m}_b(\overline{m}_b)=4196(8) ~{\rm MeV}~,
\label{eq:mb-mean}
$$
where the error comes from the most accurate determinations. 
\subsection*{\b Decay constants $f_{B_c}$ and  $f_{B_c(2S)}$}
 Using the new results in Eqs.\,\ref{eq:mb} and\,\ref{eq:mc}, we improve our previous predictions of $f_{B_c}$\,\cite{SNBc,BAGAN,SNp15} which becomes more accurate due to the inclusion of HO PT corrections and to the use of modern values of the QCD input parameters :
$$
f_{B_c}=371(17)~{\rm MeV}~(\rm Eq.~\ref{eq:fbc})
$$

An upper bound for the $B_c(2S)$ decay constant is also derived:
$$
f_{B_c}(2S)\leq (139\pm 6)~{\rm MeV}~(\rm Eq.~\ref{eq:fbc2s})~,
$$

These new results will be useful for further phenomenological analysis. 

 Improvement of our results requires a complete evaluation of the spectral function at N2LO and a future measurement of the $B_c(2S)$ decay constant. 
We plan to extend the analysis in this paper to some other $B_c$-like mesons in a future work.

\section*{References}


\begin{thebibliography}{999}
\bibitem{SN19} S. Narison, Plenary talks presented at QCD19-Montpellier-FR (2-5 july 2019) and at HEPMAD19-Antananarivo-MG (14-19 october 2019) (to appear in {\it Nucl. Part.Phys. Proc.} (2020)).
\bibitem{SVZa}M.A. Shifman, A.I. Vainshtein and V.I. Zakharov, {\it Nucl. Phys.} {\bf B147} (1979) 385.
\bibitem{SVZb}M.A. Shifman, A.I. Vainshtein and V.I. Zakharov, {\it Nucl. Phys.} {\bf B147} (1979) 448.
\bibitem{BELLa}J.S. Bell and R.A. Bertlmann, {\it Nucl. Phys.} {\bf B177}, (1981) 218.
\bibitem{BELLb}J.S. Bell and R.A. Bertlmann, {\it Nucl. Phys.} {\bf B187}, (1981) 285.
\bibitem{BECCHI}C. Becchi, S. Narison, E. de Rafael and F.J. Yndurain, {\it Z. Phys.} {\bf C8} (1981) 335.
\bibitem{SNR}S. Narison and E. de Rafael,  {\it Phys. Lett.} {\bf B103} (1981) 57.

\bibitem{ZAKA}For a review, see e.g.: V.I. Zakharov, talk given at the Sakurai's Price, {\it Int. J. Mod .Phys.} {\bf  A14}, (1999) 4865.
\bibitem{BERTa} For a review, see e.g. : R.A. Bertlmann, {\it Acta Phys. Austriaca} {\bf 53}, (1981) 305.  
\bibitem{SNB1}For a review, see e.g.: S. Narison, {\it Cambridge Monogr. Part. Phys. Nucl. Phys. Cosmol.} {\bf 17}
 (2004) 1-778 [hep-ph/0205006]. 
 \bibitem{SNB2}For a review, see e.g.: S. Narison, {\it World Sci. Lect. Notes Phys.} {\bf 26} (1989) 1.
\bibitem{SNB3}For a review, see e.g.: S. Narison, {\it Phys. Rept.}  {\bf 84} (1982) 263.
\bibitem{SNB4}For a review, see e.g.: S. Narison, {\it Acta Phys. Pol.} {\bf B 26}(1995) 687. 
\bibitem{SNREV15}For a review, see e.g.: S. Narison, {\it Nucl. Part. Phys. Proc.} {\bf 258-259} (2015) 189.
\bibitem{IOFFEb}For a review, see e.g.: B.L. Ioffe, {\it Prog. Part. Nucl. Phys.} {\bf 56} (2006) 232.
\bibitem{RRY}For a review, see e.g.: L. J. Reinders, H. Rubinstein and S. Yazaki, 
{\it Phys. Rept. } {\bf 127} (1985) 1.
\bibitem{DERAF}For a review, see e.g.: E. de Rafael, les Houches summer school, hep-ph/9802448 (1998).
\bibitem{YNDB}For a review, see e.g.: F.J Yndurain, {\it The Theory of Quark and Gluon Interactions}, 3rd edition, Springer (1999).
\bibitem{PASC} For a review, see e.g.: P. Pascual and R. Tarrach, {\it QCD: renormalization for practitioner}, Springer 1984. 
\bibitem{DOSCH} For a review, see e.g.: H.G. Dosch, {\it Non-pertubative Methods,} ed. Narison, World Scientific (1985).
\bibitem{COL}For a review, see e.g.: P. Colangelo and A. Khodjamirian, hep-ph/0010175 (2000).
\bibitem{SNBc} S. Narison, {\it Phys. Lett.} {\bf B210} (1988)  238.
 \bibitem{BAGAN}E. Bagan, H.G Dosch, P. Gosdzinsky, S. Narison and J.-M. Richard, {\it Z. Phys.} {\bf C64} (1994) 57.
 \bibitem{CHABAB}M. Chabab, {\it Phys. Lett.} {\bf B325} (1994)205.
  \bibitem{COL2}P. Colangelo, G. Nardulli and N. Paver, {\it Z. Phys.} {\bf C57} (1993) 43. 
   \bibitem{BAKER}M. Baker et al., {\it JHEP} {\bf 07} (2014) 032.
  \bibitem{SNp15}S. Narison, {\it Int. J. Mod. Phys.} {\bf A30} (2015) no.20, 1550116 and references therein.
   \bibitem{GENERALIS1}D. Broadhurst and .C. Generalis, {\it Phys. Lett.} {\bf B139} (1984) 85; ibid {\bf B165} (1985) 175. 
 \bibitem{BAGAN1} E. Bagan, J. I. Latorre, P. Pascual and R. Tarrach, {\it Nucl. Phys.} {\bf B 254} (1985) 55;  {Z. Phys.}{\bf C32}(1986) 43.
  \bibitem{GENERALIS}S.C. Generalis, Ph.D. thesis, Open Univ. report, OUT-4102-13 (1982), unpublished.
  \bibitem{FNR}E.G. Floratos, S. Narison and E. de Rafael, {\it Nucl. Phys.} 
{\bf B155} (1979) 155.

 \bibitem{CHETa}K.G. Chetyrkin and M. Steinhauser,  {\it Phys. Lett.}  {\bf B502} (2001)  104.
\bibitem{CHETb} K.G. Chetyrkin and M. Steinhauser, hep-ph/0108017. 
\bibitem{SZ}S. Narison and V.I. Zakharov, {\it Phys. Lett.} {\bf B679} (2009) 355.
\bibitem{CNZa} K.G. Chetyrkin, S. Narison and V.I. Zakharov, {\it Nucl. Phys.} 
{\bf B550} (1999)  353.
\bibitem{CNZb}S. Narison and V.I. Zakharov, {\it  Phys. Lett.} {\bf B522} (2001) 266.
\bibitem{ZAKa} For reviews, see e.g.: V.I. Zakharov, {\it Nucl. Phys. Proc. Suppl.} 
{\bf 164} (2007) 240.
\bibitem{ZAKb}S. Narison,  {\it Nucl. Phys. Proc. Suppl.} {\bf 164} 
 (2007) 225.

\bibitem{TAR}R. Tarrach, {\it Nucl. Phys.} {\bf B183} (1981) 384.

\bibitem{COQUEa} R. Coquereaux, {\it Annals of Physics} {\bf 125} (1980) 401.
\bibitem{COQUEb}P. Binetruy and T. S\"ucker, {\it Nucl. Phys.} {\bf B178} (1981) 293.

\bibitem{SNPOLEa} S. Narison, {\it  Phys. Lett.} {\bf B197} (1987) 405.
\bibitem{SNPOLEb}S. Narison, {\it  Phys. Lett.} {\bf B216} (1989) 191.

\bibitem{BROAD2a} N. Gray, D.J. Broadhurst, W. Grafe, and K. Schilcher, {\it Z. Phys.} {\bf  C48} (1990) 673.
\bibitem{BROAD2b}J. Fleischer, F. Jegerlehner, O.V. Tarasov, and O.L. Veretin, {\it Nucl. Phys.} {\bf B539}
(1999) 671.

\bibitem{CHET2a}K.G. Chetyrkin and M. Steinhauser, {\it Nucl. Phys.} {\bf B573}
(2000) 617.
\bibitem{CHET2b}K. Melnikov and T. van Ritbergen, hep-ph/9912391.
\bibitem{SNmom18} S. Narison, {\it Phys. Lett.} {\bf B784} (2018) 261.
 \bibitem{SNparam}S. Narison, {\it  Int. J. Mod. Phys.} {\bf A33} (2018) no.10, 1850045, Addendum: {\it Int. J. Mod. Phys.} {\bf A33} (2018) no.10, 1850045 and references therein.
\bibitem{SNH10a}S. Narison,  {\it Phys. Lett.} {\bf B693} (2010)  559; Erratum ibid 705 (2011) 544.
\bibitem{SNH10b}S. Narison,  {\it Phys. Lett.} {\bf B706} (2011)  412.
\bibitem{SNH10c}S. Narison,  {\it Phys. Lett.} {\bf B707} (2012)  259. 

\bibitem{PDG}M. Tanabashi et al. (Particle Data Group),{\it Phys. Rev.} {\bf D 98} (2018) 030001 and 2019 update.
\bibitem{BIJNENS}  J. Bijnens, J. Prades  and E. de Rafael, {\it Phys. Lett.} {\bf B348} (1995) 226.
\bibitem{SHIF} M.A. Shifman, arXiv:hep-ph/0009131. 
\bibitem{PERIS}  O. Cat\`a, M. Golterman and S. Peris, {\it Phys. Rev.} {\bf D77}, 093006 (2008)
\bibitem{SNTAU}  S. Narison, {\it Phys.Lett.} {\bf B673} (2009) 30.
 \bibitem{PICH}  A. Pich and A. Rodriguez-Sanchez, {\it Phys. Rev.} {\bf D94}, 034027 (2016).
\bibitem{CMS} The CMS collaboration, {\it Phys. Rev. Lett.} {\bf122} (2019)132001.
\bibitem{SNHeavy} S. Narison, {\it Phys. Lett.} {\bf B387} (1996) 162.
 \bibitem{SNHeavy2}S. Narison, {\it Nucl. Phys. (Proc. Suppl)} {\bf A54} (1997) 238.
\bibitem{SNcb3}S. Narison,  {\it Phys. Lett.} {\bf B707}  (2012) 259.
\bibitem{SNHQET13} S. Narison, {\it Phys. Lett.} {\bf B721} (2013) 269.
\bibitem{BERTb} R.A. Bertlmann,{\it Nucl. Phys.} {\bf B204}, (1982) 387.
\bibitem{BERTc} R.A. Bertlmann,{\it Non-pertubative Methods,} ed. Narison, WSC (1985). 
\bibitem{BERTd} R.A. Bertlmann,Nucl. Phys. (Proc. Suppl.) {\bf B23} (1991) 307.
\bibitem {NEUF}R. A. Bertlmann and H. Neufeld, {\it Z. Phys.} {\bf C27} (1985) 437.
\bibitem{SHAW}J. Marrow, J. Parker and G. Shaw, {\it Z. Phys.}  {\bf C37} (1987) 103.

\bibitem{SNL14} S. Narison,  {\it Phys. Lett.} {\bf B738}  (2014) 346.
\bibitem{SNp13} S. Narison, {\it Phys. Lett.} {\bf B718} (2013) 1321.
\bibitem{LAUNERb}R.A. Bertlmann, G. Launer and E. de Rafael, {\it Nucl. Phys.} {\bf B250}, (1985) 61.

  






\end{thebibliography}
 \end{document}